\documentclass[aps,prl,superscriptaddress,floatfix,reprint,longbibliography]{revtex4-2}
\usepackage{amsmath,amssymb,amsfonts}
\usepackage{graphicx}
\usepackage{xcolor}
\usepackage{braket}
\usepackage{refcount}
\usepackage{tikz}
\definecolor{lime}{HTML}{A6CE39}
\DeclareRobustCommand{\orcidicon}{
\begin{tikzpicture}
\draw[lime, fill=lime] (0,0)
circle[radius=0.16]
node[white]{{\fontfamily{qag}\selectfont \tiny \.{I}D}};
\end{tikzpicture}
\hspace{-2mm}
}
\foreach \x in {A, ..., Z}{%
\expandafter\xdef\csname orcid\x\endcsname{\noexpand\href{https://orcid.org/\csname orcidauthor\x\endcsname}{\noexpand\orcidicon}}
}

\usepackage{hyperref}
\hypersetup{hidelinks,colorlinks=true,allcolors=black,pdfstartview=Fit,breaklinks=true}

\newcommand{\TT}{\mathcal{T}}
\newcommand{\bk}[1]{\langle #1\rangle}
\newcommand{\pp}{\partial}
\newcommand{\sgn}[1]{\mathrm{sgn}(#1)}

\begin{document}
\title{Robustness of Anyonic Time-Domain Interference Against Dephasing}

\author{Hai-Dong Li}
\thanks{The authors contribute equally.}
\affiliation{National Laboratory of Solid State Microstructures, School of Physics,
 and Collaborative Innovation Center of Advanced Microstructures,
 Nanjing University, Nanjing 210093, China}

\author{Hao-Xiang Long}
\thanks{The authors contribute equally.}
\affiliation{National Laboratory of Solid State Microstructures, School of Physics,
 and Collaborative Innovation Center of Advanced Microstructures,
 Nanjing University, Nanjing 210093, China}

\author{Gu Zhang}
\affiliation{National Laboratory of Solid State Microstructures, School of Physics,
 and Collaborative Innovation Center of Advanced Microstructures,
 Nanjing University, Nanjing 210093, China}

\author{D. Y. Xing}
\affiliation{National Laboratory of Solid State Microstructures, School of Physics,
 and Collaborative Innovation Center of Advanced Microstructures,
 Nanjing University, Nanjing 210093, China}

\author{Wei Chen \hspace{-1.5mm}\orcidA{}}
\email{Corresponding author: pchenweis@gmail.com}
\affiliation{National Laboratory of Solid State Microstructures, School of Physics,
 and Collaborative Innovation Center of Advanced Microstructures,
 Nanjing University, Nanjing 210093, China}
\affiliation{Jiangsu Physical Science Research Center and Jiangsu Key Laboratory of Quantum Information Science and Technology, Nanjing University, Nanjing 210093, China}

\begin{abstract}
 Recent anyon-collider experiments have identified time-domain braiding (TDB) as a new route to probing anyonic statistics. However, how such time-domain interference responds to environmental noise remains an open question. Here we show that TDB interference is intrinsically more robust than its spatial counterpart. In conventional spatial interference, environmental noise induces phase fluctuations that accumulate throughout all paths, leading to exponential suppression of the interference as the dephasing rate increases. In TDB, by contrast, the interfering temporal paths experience largely overlapping environmental fluctuations, whose contributions cancel out in the relative phase, so that only a small nonoverlapping temporal window contributes to dephasing. As a result, TDB interference decays only algebraically with the dephasing rate, providing a new experimental hallmark for its identification. We propose a Fabry-Pérot fractional quantum Hall setup in which both types of interference coexist and experience the same environmental noise, enabling a direct comparison of their dephasing behaviors. Our results establish robustness against dephasing as a fundamental advantage of TDB and suggest new opportunities for exploiting anyonic interference in noisy mesoscopic systems.
\end{abstract}

\maketitle

\emph{Introduction.---}Quantum interference, rooted in the coherent superposition principle, lies at the heart of quantum mechanics~\cite{sakuraiModernQuantumMechanics2020, feynmanQuantumMechanicsPath2010}.
For a free particle, interference provides a direct manifestation of the phase information accumulated along different Feynman paths, and has therefore become a central paradigm for probing phase coherence~\cite{jaklevicQuantumInterferenceEffects1964,colellaObservationGravitationallyInduced1975,webbObservationAharonovBohmOscillations1985,kasevichAtomicInterferometryUsing1991,andrewsObservationInterferenceTwo1997}, topological properties~\cite{pengAharonovBohmInterference2010, chen16prl, chen19prl}, and quantum statistics~\cite{brownCorrelationPhotonsTwo1956, hennyFermionicHanburyBrown1999, jeltesComparisonHanburyBrown2007, kimblePhotonAntibunchingResonance1977}.
Mesoscopic electronic systems offer an ideal platform for exploring a wide variety of quantum interference effects, as electron phase coherence can persist over the characteristic length scales of the device~\cite{beenakkerQuantumTransportSemiconductor1991, dattaElectronicTransportMesoscopic1995, imryIntroductionMesoscopicPhysics1998, washburnAharonovBohmEffectNormal1986}, while interferometric geometries can be engineered with high precision.
Representative examples include Aharonov-Bohm~\cite{aharonovSignificanceElectromagneticPotentials1959, washburnAharonovBohmEffectNormal1986}, Fabry-Pérot (FP)~\cite{liangFabryPerotInterference2001}, and Mach-Zehnder (MZ)~\cite{jiElectronicMachZehnder2003} interferometers. 
Despite their diverse physical realizations, these phenomena share a common underlying feature: interference occurs between distinct propagation paths in real space.
This spatial interference paradigm has enabled the detection of anyonic braiding statistics in recent FP~\cite{nakamuraDirectObservationAnyonic2020, nakamuraImpactBulkedgeCoupling2022, nakamuraFabryPerotInterferometry22023, willettInterferenceMeasurementsNonAbelian2023,KimNN2024, werkmeisterStronglyCoupledEdge2024, werkmeisterAnyonBraidingTelegraph2025,KimNature2026,KimX26} and MZ~\cite{HeiblumNP2023,GhoshNP25} experiments, supported by theoretical advances~\cite{LawPRB06,FeldmanKitaevPRL06,halperinTheoryFabryPerotQuantum2011}.

\begin{figure}[htbp]
 \centering
 \includegraphics[width=0.45\textwidth]{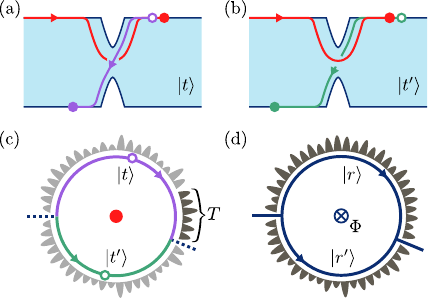}
 \caption{
(a,b) Schematic illustration of the two interfering tunneling events at the QPC, \(|t\rangle\) and \(|t'\rangle\), respectively. The tunneling event creates a quasiparticle (filled circle) and a quasihole (open circle), occurring (a) after and (b) before the arrival of the injected nonequilibrium anyon (red filled circle). 
(c) Schematic illustration of time-domain interference in the presence of environmental noise. The two paths correspond to the tunneling events in (a) and (b). The light-gray regions denote the common noise history shared by the two paths, while the dark-gray region of duration \(T=t-t'\) denotes the remaining non-overlapping noise history that contributes to dephasing.
(d) Schematic illustration of spatial interference between two paths, $|r\rangle$ and $|r'\rangle$, controlled by the magnetic flux \(\Phi\). The dark-gray regions denote the uncorrelated environmental noise accumulated along the two paths.
 }
 \label{fig:example01}
\end{figure}

In parallel with this spatial route, recent theoretical and experimental progress in anyon colliders has revealed a qualitatively different paradigm for probing anyonic braiding statistics~
\cite{rosenowCurrentCorrelationsMesoscopic2016, hanTopologicalVacuumBubbles2016, leeNegativeExcessShot2019, leeFractionalMutualStatistics2020, leeNonAbelianAnyonCollider2022, bartolomeiFractionalStatisticsAnyon2020, mora2022anyonic, morelFractionalizationAnyonicStatistics2022,  glidicCrossCorrelationInvestigationAnyon2023, jonckheereAnyonicStatisticsRevealed2023, leePartitioningDilutedAnyons2023, schillerAnyonStatisticsConductance2023, zhangTunnelingCurrentCurrent2024,han2024PRL,ThammBerndPRL24,IyerPRL2024, ruelleTimedomainBraidingAnyons2025,DeNC25, zhangLandscapesOutofEquilibriumAnyonic2025,AlkalayX2026}.
Unlike FP and MZ interferometry, where braiding is tied to spatial winding, the collider mechanism is interpreted as time-domain braiding (TDB)~\cite{schillerAnyonStatisticsConductance2023, hanTopologicalVacuumBubbles2016,leeNegativeExcessShot2019,leeFractionalMutualStatistics2020,leeNonAbelianAnyonCollider2022}: 
an interference effect between tunneling histories that differ in their temporal ordering. Specifically, a quasiparticle-quasihole pair (the filled-open pair of circles in Fig.~\ref{fig:example01}) can be generated at the quantum point contact (QPC) either after [Fig.~\ref{fig:example01}(a)] or before [Fig.~\ref{fig:example01}(b)] the arrival of injected nonequilibrium anyons (the red filled circle), giving rise to interference between two temporal paths involving an effective anyon braiding [Fig.~\ref{fig:example01}(c)]. As TDB is fundamentally a quantum interference phenomenon, a natural question arises: How does it respond to environmental dephasing? Addressing this question is essential for understanding the distinctive nature of TDB and assessing its potential applications in realistic quantum systems.

\begin{figure}[tbp]
 \centering
 \includegraphics[width=0.45\textwidth]{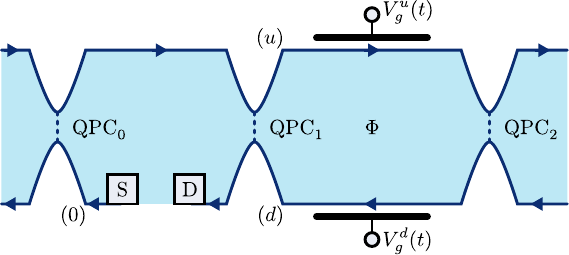}
\caption{
FP interferometer formed by fractional quantum Hall edge states. Anyons emitted from source S propagate along edge \((0)\) and tunnel through QPC$_0$ to form a dilute beam. QPC$_1$ and QPC$_2$ couple the upper \((u)\) and lower \((d)\) edges, enclosing a magnetic flux \(\Phi\). The fluctuating gates \(V_g^u\) and \(V_g^d\) model environmental noise, and the output current is measured at drain D.
}
 \label{fig:example02}
\end{figure}

In this Letter, we investigate TDB in the presence of noise-induced dephasing and demonstrate that TDB interference is intrinsically more robust against dephasing than spatial interference. In contrast to spatial interference, where the dephasing effect accumulates throughout all paths [Fig.~\ref{fig:example01}(d)], the interfering temporal histories in TDB experience largely overlapping environmental fluctuations [light-gray noise shading in Fig.~\ref{fig:example01}(c)]. These shared fluctuations cancel out in the relative phase, leaving only a small nonoverlapping temporal window susceptible to dephasing [dark-gray noise shading in Fig.~\ref{fig:example01}(c)]. Consequently, the environmental noise leads to a \textit{power-law} suppression of the TDB interference, distinct from the \textit{exponential} suppression in spatial interference. Such power-law behavior under dephasing provides a distinctive experimental hallmark of TDB. We further propose a fractional quantum Hall FP setup (cf. Fig.~\ref{fig:example02}) in which TDB and spatial interference coexist and experience the same environmental noise, providing a direct comparison between distinct dephasing behaviors.

\emph{Distinct dephasing mechanisms.---}To highlight the  distinctive dephasing behavior of TDB, it is useful to first recall the basic dephasing picture in conventional spatial interference.
Consider the typical two-path interferometer illustrated in Fig.~1(d), where an incoming particle can propagate along two paths, denoted by \(|r\rangle\) and \(|r'\rangle\), with the interference phase continuously tuned by the enclosed magnetic flux \(\Phi\).
After a certain propagation time, the two coherent paths recombine at the right terminal, and the interference term can be written as
\begin{equation}\label{rr}
 A_S(r,r')=\langle r|r'\rangle
 =
 e^{i\theta_{{B}}}
 e^{i[\theta_2(r')-\theta_1(r)]}
 \langle r|r'\rangle_{0},
\end{equation}
where \(\theta_{{B}}=2\pi\Phi/\Phi_0\) is the Aharonov-Bohm phase, \(\Phi_0\) is the flux quantum, \(\theta_1\) and \(\theta_2\) are the random phases, induced by environmental fluctuations, along the paths \(|r\rangle\) and \(|r'\rangle\), respectively, and \(\langle r|r'\rangle_0\) denotes the interference term in the absence of the magnetic flux and noise.

Dephasing arises after averaging the phase factor in Eq.~\eqref{rr} over environmental noise~\cite{sternPhaseUncertaintyLoss1990}.
When the accumulated phases arise from many independent fluctuation events,
the central-limit theorem justifies a Gaussian approximation, leading to
$\langle
 e^{i(\theta_2-\theta_1)}
 \rangle_{\xi}
 =
 e^{
   -\frac{1}{2}
   \langle
   (\theta_2-\theta_1)^2
   \rangle_{\xi}
  }$, 
where \(\langle \cdots\rangle_\xi\) denotes averaging over the environmental noise ensemble \(\xi\). The mean phase difference is set to zero,
\(\langle{\theta_2-\theta_1}\rangle_{\xi}=0\), without loss of generality.  We further assume that the noise is short-range correlated
along each path and uncorrelated between different paths. The former
makes the phase variance grow linearly with the path length, while the
latter eliminates the cross correlation between the two paths. Thus,
$\frac{1}{2}\left\langle
 (\theta_2-\theta_1)^2\right\rangle_\xi=\frac{1}{2}(\langle
 \theta_2^2\rangle_\xi+\langle
 \theta_1^2\rangle_\xi)\simeq (r+r')/l_\phi$,
where \(r\) and \(r'\) are the lengths of the two paths, \(l_\phi\)
is the phase coherence length. The averaged interference term then becomes
\begin{equation}\label{der}
 \langle A_S(r,r') \rangle_\xi
 =e^{-(r+r')/l_\phi}
 e^{i\theta_{{B}}}
 \langle r|r'\rangle_{0},
\end{equation}
in which the decay factor reflects the additive dephasing from
uncorrelated fluctuations along the two spatial paths.

This picture changes qualitatively in TDB, where the interference occurs between temporal paths. As
illustrated in Figs.~\ref{fig:example01}(a,b), these two paths, denoted by \(|t\rangle\) and \(|t'\rangle\), correspond to anyonic
tunneling events (or equivalently, creation of quasiparticle-quasihole pairs) that occur at the QPC after or before the arrival of the injected nonequilibrium anyons, respectively.
This process can be interpreted as TDB: the
quasihole follows two temporal paths that braid around the injected anyon~\cite{schillerAnyonStatisticsConductance2023, ruelleTimedomainBraidingAnyons2025,zhangLandscapesOutofEquilibriumAnyonic2025}. For a Laughlin state with filling
fraction \(\nu=1/(2n+1)\), with \(n\) a positive integer, the TDB interference term in the presence of noise can be written as
\begin{equation}\label{tt}
 A_T(t,t')= \langle t|t'\rangle
 = e^{i\vartheta_A}
 e^{i[\vartheta(t')-\vartheta(t)]}
 \langle t|t'\rangle_0,
\end{equation}
where \(\vartheta_A=2\pi \nu\) is the braiding phase between a single pair of anyons,
\(\vartheta(t)\) and \(\vartheta(t')\) are the random phases induced by
environmental noise along the temporal paths \(|t\rangle\) and
\(|t'\rangle\), respectively, and
$
 \langle t|t'\rangle_0
$
is the equilibrium interference term to be
specified below.

By comparing Eqs.~\eqref{rr} and~\eqref{tt}, one sees that the TDB
interference term has a form analogous to its spatial counterpart. The key
difference, however, lies in the correlation structure of the noise-induced phases.
In spatial interference, the random phases \(\theta_1\) and \(\theta_2\) are generally uncorrelated because the two interfering paths experience independent environmental fluctuations. In TDB, by contrast, the phases \(\vartheta(t)\)
and \(\vartheta(t')\) are generated by the same noise history, sampled over two different temporal intervals. The two tunneling events are separated by the relative time \(T\equiv t-t'\). As a result, the phase contributions from the overlapping portion of the evolution cancel each other, leaving only a non-canceling time window \(|T|\), as illustrated in
Fig.~\ref{fig:example01}(c). Assuming Gaussian phase fluctuations with short-range correlations, the noise-averaged TDB interference term becomes
\begin{equation}\label{eq:at}
 \langle A_T(t,t')\rangle_\xi
 = e^{-|T|/\tau_{\phi}}e^{i\vartheta_A}
 \langle t|t'\rangle_0,
\end{equation}
where $\tau_\phi$ is the phase coherence time. 

In comparison to spatial dephasing, where the correlation is suppressed by the entire length of the loop [cf. $e^{-(r+r')/l_\phi}$ of Eq.~\eqref{der}], Eq.~\eqref{eq:at} indicates, importantly, that TDB is influenced only by dephasing within the temporal window, $|T|$. The latter influence is thus much smaller than the spatial one. In the limit \(T\to0\), the noncanceling window vanishes, and no dephasing takes place~\footnote{\label{fn:tau}Here, $T\to0$ does not imply that the two times become arbitrarily close; one must still require $|T|$ to remain larger than the temporal broadening $1/(e^* V)$ of the nonequilibrium anyon.}. Physically, this robustness stems from the fact that the two interfering histories share nearly the same environmental record, making them difficult to distinguish.

For a single nonequilibrium anyon arriving at the QPC at \(t_0\), the
tunneling current just after its passage, \(I_T(t)\) with $t=t_0^+$, is
evaluated by summing the TDB interference between the tunneling event at
\(t\) and all earlier events with \(t'<t_0\)~\cite{sm}:
\begin{equation}
I_T(t) \propto \int_{-\infty}^{t_0} dt'\,
 \left|\langle A_T(t, t')\rangle_\xi \right| \propto \left(\frac{1}{\tau_{\phi}}\right)^{2\nu-1}.
 \label{eq:oneanyoncurr}
\end{equation}
The algebraic dependence of the interference current on the dephasing rate \(1/\tau_\phi\) is in
stark contrast to the exponential suppression in spatial
interference.

\emph{Setup and theoretical framework.---}We now demonstrate this contrast in a concrete
setup where the two types of interference coexist and are exposed to the
same environmental noise. Specifically, consider the fractional quantum Hall FP geometry shown in Fig.~\ref{fig:example02}.
The system contains three edge channels \(s=0,u,d\), denoting the injection, upper, and lower edges, respectively.
QPC$_0$ at \(x_0\) dilutely injects anyons into the upper edge, whereas QPC$_1$ and QPC$_2$ at \(x_1\) and \(x_2\) couple the upper and lower edges to form the FP interferometer.
This geometry supports local TDB processes at QPC$_1$ and QPC$_2$, together with spatial interference between them. We use the standard chiral Luttinger liquid description of the edge states and introduce weak quasiparticle tunneling at QPCs~\cite{wen1990PRB,KaneFisherPRL94,FendleyLudwigSaleurPRB95}.

The edge channel $s$ is described by $H_{s}  = \frac{v_F}{4\pi} \int dx\, (\partial_x \phi_{s})^2$, where \(\phi_s(x)\) is the chiral bosonic field propagating with velocity
\(v_F\) and satisfying
$
 [\phi_s(x),\phi_{s'}(x')]
 =\chi_s i\pi \sgn{x-x'}\delta_{ss'}.
$
Here, $\chi_u=+1$ and $\chi_d=-1$ characterize the opposite chiralities of the two edges.
Environmental noise is
modeled by fluctuating gate potentials \(V_g^s(x,t)\) acting on the
interferometer edges (see Fig.~\ref{fig:example02}), with $H^s_g= \chi_s\frac{\sqrt{\nu}}{2\pi} \int dx \left[  e V_{g}^{s}(x, t) \partial_x \phi_{s} \right]$. 

Taking \(H_0=\sum_{s=0,u,d}H_s+\sum_{s=u,d}H_g^s\) as the unperturbed
Hamiltonian, we treat quasiparticle tunneling at the QPCs as a perturbation. In the interaction picture, the tunneling Hamiltonian is $H_T(t)=\sum_{j=0,1,2}[\gamma_j \TT_j(t)+\text{h.c.}]$, where
\begin{equation}
 \begin{split}
  \TT_0(t) & =
  e^{-i\vartheta_0(t)}
  \psi_{u}^\dagger(x_{0},t) \psi_{0}(x_{0},t),                     \\
  \TT_1(t) & =
  e^{i\vartheta_1(t)}
  \psi_d^\dagger(x_{1},t) \psi_u(x_{1},t),                         \\
  \TT_2(t) & =
  e^{i[\theta_B^*-\vartheta_2(t)]}
  \psi_d^\dagger(x_{2},t) \psi_u(x_{2},t).
 \end{split}
\end{equation}
Here \(\gamma_j\) is the tunneling amplitude at QPC\(_j\), chosen to be
real without loss of generality. 
The anyon creation operator on edge \(s\) reads \(\psi_s^\dagger(x,t)\sim e^{-i\sqrt{\nu}\phi_s(x,t)}\).

We specify the phase factors appearing in the tunneling operators.
With a convenient gauge choice, the dc bias voltage \(V\) applied to channel \(0\) that injects anyons is incorporated into \(\TT_0\) through the phase
\(\vartheta_0(t)=e^*Vt\). The Aharonov-Bohm phase
\(\theta_B^*=2\pi\Phi/\Phi_0^*\) is attached to \(\TT_2\), where \(\Phi_0^*=h/e^*\) is
the flux quantum associated with the fractional charge \(e^*=\nu e\).
The fluctuating gate potentials \(V_g^s(x,t)\)
induce phases in the tunneling operators \(\TT_{j=1,2}\)~\cite{rech2017PRL},
\begin{equation}
     \vartheta_{j}(t)
     =
     e^* \int_{-\infty}^t d\tau\,
     V_g^{s}\bigl(x_{j} -\chi_{s(j)} v_F(t-\tau),\tau\bigr),
\end{equation}
where $s(1)=d$ and $s(2)=u$ specify the edge
associated with each QPC.


The interference effect is manifested by the current measured at the drain, with the current operator
$
    \hat{I}(t)
     =
     -\frac{ev_F\sqrt{\nu}}{2\pi}
 \pp_x\phi_d(x,t).
$
Its expectation value can be evaluated on the Keldysh time contour~\cite{keldysh1964diagram,chamon1995PRB,chamon1996PRB,rammer1986RMP},
\begin{equation}
I(t)
 =
 \frac{1}{2}\sum_{\eta=\pm} \left<
  \hat{T}_K \hat{I}(t_{\eta})
  e^{-i\int_{K} dt'\, H_T(t')}
 \right>_0,
\end{equation}
where \(\hat{T}_K\) is the Keldysh time-ordering operator, \(\eta\) is the Keldysh branch index and \(\bk{\cdots}_{0}\) denotes an expectation value evaluated in equilibrium.
Keeping the leading order in the tunneling amplitudes $\gamma_{1,2}$, the total current separates into
$
 {I}
 =
 {I_{\text{TDB}}}
 +
 {I_{\text{FP}}}.
$
The first term, \({I_{\text{TDB}}}\propto \gamma_1^2, \gamma_2^2\), comes from tunneling at QPC$_1$ and QPC$_2$ induced by local TDB process, whereas the second term \({I_{\text{FP}}}\propto \gamma_1\gamma_2\) arises from the FP interference between the two QPCs that can be tuned by \(\theta_B^*\)~\footnote{Here, the FP interference between two QPCs is induced by a \emph{nonlocal} TDB mechanism. Nevertheless, its physical manifestation remains the familiar spatial interference~\cite{hanTopologicalVacuumBubbles2016,girdhar2026arxiv,han2024PRL}, in sharp contrast to the local TDB effect.}.
Further averaging over the noise yields~\cite{sm}
\begin{align}
  \left\langle I_{\text{TDB}}(t)\right\rangle_\xi
    &=
  e^* \sum_{\eta=\pm}\sum_{j=1,2} \gamma_j^2\eta
      \int dt'\, \left\langle  A_{T,\eta}^{jj}(t,t')\right\rangle_\xi, \label{II_TDB}\\
  \left\langle{I_{\text{FP}}(t)}\right\rangle_\xi
    &=
  e^* \sum_{\eta=\pm}\sum_{j = 1,2} \gamma_j\gamma_{\bar{j}}\eta
  \int dt'\,
   \left\langle A_{T,\eta}^{j\bar{j}}(t,t')\right\rangle_\xi,\label{II_FP}
\end{align}
where \(\bar{j}\) is complementary to \(j\).
The interference term is
$
A_{T,\eta}^{jj'}(t,t')
=
e^{i\vartheta_A^{jj'}(T)}e^{-i(j-j')\theta_B^*}
$
$
\times e^{-i[p_j\vartheta_j(t)-p_{j'}\vartheta_{j'}(t')]}
\langle t|t'\rangle_{\eta}^{jj'} ,
$ where 
\(\langle t|t'\rangle_{\eta}^{jj'}
=
a^{2\nu-2}/\{(2\pi)^2 [a+i\eta (v_F T-L_{jj'})]^{\nu} [a+i\eta (v_F T+L_{jj'})]^{\nu}\}
\) is the equilibrium interference term with $L_{jj'} = x_j-x_{j'}$ and $a$ the short-distance cutoff, $p_j=(-1)^{j+1}$, and
$\vartheta_A^{jj'}$ takes the form, in the dilute-injection regime~\cite{rosenowCurrentCorrelationsMesoscopic2016,girdhar2026arxiv,schillerAnyonStatisticsConductance2023}
\begin{equation}
     \vartheta_A^{jj'}(T)
     =
     i\frac{|T-L_{jj'}/v_F|}{\tau_{\delta}}
     \left(
      1-e^{i2\pi\nu\cdot\text{sgn}(T-L_{jj'}/v_F)}
     \right).
\end{equation}
Here, \(\tau_\delta=e^*/I_{0}\) is the mean difference in time between two neighbouring injected anyons, with \(I_{0}\) the injection current at QPC\(_0\). For $j=j'$,
$A_{T,\pm}^{jj}$ reduces to the form of Eq.~\eqref{tt}, where $j$ labels the two QPCs and the corresponding TDB phase is \(\vartheta_A^{jj}\).

\emph{Dephasing of FP and TDB interference.---}The dephasing effect is modeled by the random gates, whose gate potential is assumed to be spatially uniform over a region of length \(l_g\),
\(V_g^s(x,t)=W_s(x)V_s(t)\), where $W_s(x)=1$ for $|x-x_g|<l_g/2$ and $W_s(x)=0$ otherwise. Here, $x_g$ denotes the gate center and \(V_s(t)\) is the fluctuating gate voltage on edge \(s\). Such random gates may either be implemented intentionally to induce dephasing or simulate realistic environmental charge fluctuations that couple capacitively to the edges~\cite{marquardtInfluenceDephasingShot2004,
roulleauNoiseDephasingEdge2008,halperinTheoryFabryPerotQuantum2011,
hanTopologicalVacuumBubbles2016}. 
We adopt a zero-mean Gaussian noise for \(V_s(t)\) with zero correlation between different edges and short-time correlation within the same edge, meaning that
\begin{equation}
     \bk{V_s(\tau_1)V_{s'}(\tau_2)}_{\xi}
     =\delta_{s,s'}
     \frac{f_{\xi}(\tau_1-\tau_2)}{\nu^2 e^2\tau_\phi}.
\end{equation}
Here, \(f_{\xi}\) is the temporal correlation function, and \(\tau_\phi\) measures the noise strength.

The currents~\eqref{II_TDB} and \eqref{II_FP} are obtained by averaging over the noise history. 
We denote the noise-averaged phase correlators by \(\mathcal K_{\mathrm{TDB}}^{\xi}(j)=\left\langle
 e^{-i[p_j \vartheta_j(t)-p_j\vartheta_j(t')]}
 \right\rangle_{\mathrm{\xi}}\) and \(\mathcal K_{\mathrm{FP}}^{\xi}(j)=\left\langle
 e^{i[p_j \vartheta_j(t)-p_{\bar j}\vartheta_{\bar j}(t')]}
 \right\rangle_{\mathrm{\xi}}\) for local TDB and FP interference, respectively.
In the white-noise limit \(
f_{w}(\tau)=\delta(\tau)
\), which applies when the environmental correlation time is much shorter than the characteristic time scale of anyon dynamics, we obtain~\cite{sm}
\begin{align}
 \mathcal K_{\mathrm{TDB}}^{w}
 =e^{-\min(|T|,\tau_g)/\tau_\phi},\ \ \ 
 \mathcal K_{\mathrm{FP}}^{w}
 =e^{-\tau_g/\tau_\phi},
 \label{eq:WN_average}
\end{align}
where \(\tau_g \equiv l_g/v_F\) is the traversal time through the noisy gate region. Equation~\eqref{eq:WN_average} reveals a clear contrast between local TDB and FP interference for $|T|<\tau_g$.

To characterize the dephasing effect, we define the normalized current as the ratio of the current with dephasing to that without dephasing. For the FP contribution, the dephasing factor in Eq.~\eqref{eq:WN_average} is independent of the relative time and can therefore be factored out of the integral in Eq.~\eqref{II_FP}, yielding
\begin{equation}
 D_{\mathrm{FP}}(\tau_\phi)
 \equiv
 \frac{\bk{I_{\mathrm{FP}}}_{w}(\tau_\phi)}
 {\bk{I_{\mathrm{FP}}}_{w}(\infty)}
 =e^{-\tau_g/\tau_\phi},
 \label{eq:FP_normalized_definition}
\end{equation}
which exhibits conventional exponential suppression as noise strength increases~\cite{sm}.

For TDB, the dephasing effect decreases continuously as the two tunneling events
approach one another in time. In particular, as \(|T|\to0\),
\(\mathcal K_{\mathrm{TDB}}^{w}\to1\), since the two interfering paths experience the same noise history and dephasing consequently vanishes. Noise averaging Eq.~\eqref{II_TDB} and substituting $\mathcal K_{\mathrm{TDB}}^{w}$ in
Eq.~\eqref{eq:WN_average} gives~\cite{sm}
\begin{gather}\label{ITDB}
 \bk{I_{\mathrm{TDB}}}_{w}(\tau_\phi)
 =\frac{(\tau_g/\tau_\delta)^{1-2\nu}}{\Gamma(1-2\nu)}
 \mathrm{Im}\!\left\{\mathcal F\right\}\mathcal I_2, \\
 \mathcal I_2
 =\frac{e^*(\gamma_1^2+\gamma_2^2)}{\pi^2}
 \frac{\tau_\delta^{1-2\nu}}{v_F^{2\nu} a^{2-2\nu}}
 \Gamma(1-2\nu)\sin(\pi\nu),  \nonumber  \\
 \begin{aligned}
 \mathcal F
 ={}&\left(z_0+\frac{\tau_g}{\tau_\phi}\right)^{2\nu-1} 
 \left[\Gamma(1-2\nu)-\Gamma\!\left(1-2\nu,z_0+\frac{\tau_g}{\tau_\phi}\right)\right] \\
 &+e^{-\tau_g/\tau_\phi}z_0^{2\nu-1}\Gamma(1-2\nu,z_0),
 \end{aligned} \nonumber
\end{gather}
where \(z_0=(\tau_g/\tau_\delta)(1-e^{i2\pi\nu})\), $\Gamma(u)$ and $\Gamma(u,z)$ denote the Euler
and the upper incomplete Gamma function, respectively.
The normalized TDB current is defined as
\begin{equation}
 D_{\mathrm{TDB}}(\tau_\phi)
 \equiv
 \frac{\bk{I_{\mathrm{TDB}}}_{w}(\tau_\phi)}
 {\bk{I_{\mathrm{TDB}}}_{w}(\infty)}.
 \label{eq:TDB_normalized_definition}
\end{equation}

\begin{figure}[t]
 \centering
 \includegraphics[width=0.9\columnwidth]
{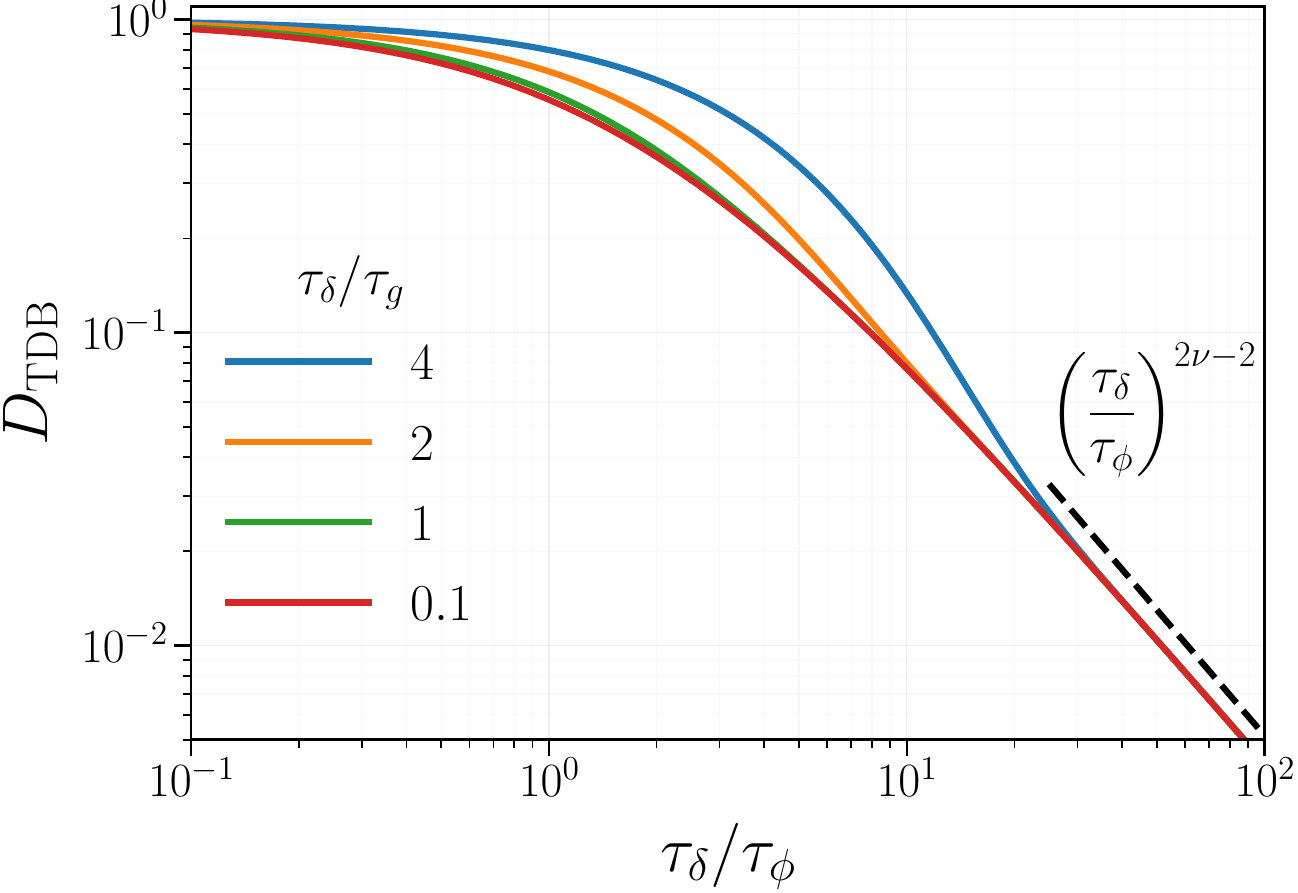}
 \caption{
  Normalized TDB current \(D_{\mathrm{TDB}}\) as a function of \(\tau_\delta/\tau_\phi\) for \(\nu=1/3\) and different values of \(\tau_\delta/\tau_g\). The dashed line is a guide to the asymptotic power-law scaling \(D_{\mathrm{TDB}}\propto(\tau_\delta/\tau_\phi)^{2\nu-2}\).
 }
 \label{fig:dephasing}
\end{figure}

We analyze the expression in two dephasing limits. For dilute anyon injection, the temporal interference is confined to a time window of order \(\tau_\delta\)~\cite{schillerAnyonStatisticsConductance2023,leePartitioningDilutedAnyons2023,leeNonAbelianAnyonCollider2022}. As a result, for \(\tau_\delta/\tau_\phi\ll1\), the dephasing effect becomes negligible and \(D_{\mathrm{TDB}}\to1\). In the strong-dephasing regime with
\(
 \tau_\delta,\tau_g\gg \tau_\phi\gg (e^*V)^{-1},
\)
the TDB contribution is governed by the dephasing time scale \(\tau_\phi\). In this limit,
Eq.~\eqref{eq:TDB_normalized_definition} can be simplified to~\cite{sm}
\begin{equation}\label{D}
 D_{\mathrm{TDB}}
 \simeq C_\nu
 \left(\frac{\tau_\delta}{\tau_\phi}\right)^{2\nu-2},
 \quad
 C_\nu=
 \frac{(1-2\nu)\sin(2\pi\nu)}
 {\mathrm{Im}[(1-e^{i2\pi\nu})^{2\nu-1}]},
\end{equation}
revealing a power-law decay with increasing dephasing rate. As shown in Fig.~\ref{fig:dephasing}, the numerical results for different \(\tau_\delta\)  at fixed $\tau_g$ asymptotically converge to a common straight line with slope \(2\nu-2\) on the log-log scale, in agreement with Eq.~\eqref{D}. Consistently, extending the result for deterministic single-anyon injection in Eq.~\eqref{eq:oneanyoncurr} to dilute Poissonian injection yields the same \((1/\tau_\phi)^{2\nu-2}\) scaling~\cite{sm}.
Compared with the exponential suppression of FP interference, the algebraic decay of TDB demonstrates its enhanced robustness against environmental noise. Equation~\eqref{D} further indicates that the dephasing can be suppressed by reducing \(\tau_\delta\), experimentally achievable by increasing the injection current \(I_0\).

Beyond the white-noise limit, we further consider environmental noise with a finite correlation time. 
As the correlation time of the environmental noise increases, dephasing of TDB interference is further reduced, since only the temporally uncorrelated part of the phase fluctuations contributes to dephasing.
In the strong-dephasing regime, even the phase fluctuations within the dephasing window become temporally correlated. Accordingly, the normalized TDB current exhibits a weaker algebraic decay,
${D}_{\mathrm{TDB}}
 \propto \left({\tau_\delta}/{\tau_\phi}\right)^{\nu-1},$
with the exponent reduced by a factor of two~\cite{sm}.

\emph{Discussion.---}The proposal can be readily implemented in existing experimental settings. The total current consists of the FP and TDB contributions, $I=I_{\mathrm{FP}}+I_{\mathrm{TDB}}$, which can be separately extracted from their distinct magnetic-flux dependence. In particular, the FP current oscillates with the magnetic flux, whereas the TDB current is flux-independent. By intentionally introducing random gates, one can then compare the dephasing behaviors of $\langle I_{\mathrm{FP}} \rangle$ and $\langle I_{\mathrm{TDB}}\rangle$, revealing their exponential and power-law suppression, respectively. The scaling law of TDB dephasing can also be independently measured in a single-QPC setup. 
The characteristic timescale $\tau_\delta$ of TDB interference can be tuned by the injection current $I_0$. Therefore, even at fixed environmental noise, the power-law scaling in the strong-dephasing regime can be directly probed by varying $I_0$, with $D_{\mathrm{TDB}}\propto\tau_\delta^{2\nu-2}\propto I_0^{2-2\nu}$ [cf. Eq.~\eqref{D}].

As mentioned earlier, our theoretical model also captures realistic environmental fluctuations that couple capacitively to anyons propagating along the edges. Let $\lambda$ denote the characteristic length scale of the spatial charge fluctuations, playing the role of $l_g$ in the random-gate model. Assuming a Markovian environment, an anyon traversing one arm of the interferometer experiences $N=L_{21}/\lambda$ independent dephasing events. In the strong-dephasing limit, the resulting TDB dephasing factor becomes $D_{\mathrm{TDB}}\simeq C_\nu \left(N\tau_\delta/\tau_\phi\right)^{2\nu-2}$. Thus, even after multiple independent dephasing events, TDB retains the algebraic suppression~\cite{sm}. In contrast, for FP interference, successive dephasing effects accumulate exponentially, yielding 
$D_{\mathrm{FP}}=\exp[-N\lambda/(v_F\tau_\phi)]=\exp[-L_{21}/(v_F\tau_\phi)]$.
This qualitative distinction persists in realistic fluctuating environments, pointing to a general robustness of the TDB interference against environmental dephasing. Our results may stimulate further exploration of robust anyonic phenomena in noisy quantum environments.

\begin{acknowledgments}
 This work was supported by the National Natural Science Foundation of China (No. 12574051), the Natural Science Foundation of Jiangsu Province (Nos. BK20250008, BK20233001, BK20253009), the Fundamental Research Funds for the Central Universities (Nos. 2026010801, KG202501), and the National Key Projects for Research and Development of China (No. 2022YFA1204701).
\end{acknowledgments}

\bibliography{refsMain}

\end{document}